\documentclass[aps,pre,twocolumn,showpacs,floatfix]{revtex4} 
\usepackage{amsmath,amssymb,isolatin1,graphicx,times}

\newcommand{\be}{\begin{equation}}
\newcommand{\ee}{\end{equation}}
\newcommand{\bea}{\begin{eqnarray}}
\newcommand{\eea}{\end{eqnarray}}
\newcommand{\bw}{\begin{widetext}}
\newcommand{\ew}{\end{widetext}}
\newcommand{\ba}{\beta}

\newcommand{\n}{\nu} 
\newcommand{\rd}{{\rm d}}
\newcommand{\vvr}{\vec{r}} 
\newcommand{\vvv}{\vec{v}}
\newcommand{\vve}{\vec{E}}
\newcommand{\cn}{{\cal{N}}}
\newcommand{\rhom}{\boldsymbol{\rho}}

\newcommand{\half}{\frac 1 2}
\newcommand{\arccot}{\,{\rm arccot}}

\begin{document}
 
\title{Thermodynamic formalism for field driven Lorentz gases}
\author{Oliver M{\"u}lken}
\email{O.Mulken@phys.uu.nl}
\author{Henk van Beijeren}
\email{H.vanBeijeren@phys.uu.nl}
\affiliation{Institute for Theoretical Physics, Utrecht University,
Leuvenlaan 4, 3584 CE Utrecht, The Netherlands}
 
\date{\today} 
\begin{abstract}
We analytically determine the dynamical properties of two dimensional
field driven Lorentz gases within the thermodynamic formalism.  For dilute
gases subjected to an iso-kinetic thermostat, we calculate the topological
pressure as a function of a temperature-like parameter $\ba$ up to second
order in the strength of the applied field.  The Kolmogorov-Sinai entropy
and the topological entropy can be extracted from a dynamical entropy
defined as a Legendre transform of the topological pressure. Our
calculations of the Kolmogorov-Sinai entropy exactly agree with previous
calculations based on a Lorentz-Boltzmann equation approach.  We give
analytic results for the topological entropy and calculate the dimension
spectrum from the dynamical entropy function. 
\end{abstract}
\maketitle

\section{Introduction}

In dynamical system theory the Lorentz gas acts as a paradigm allowing to
address fundamental issues of non-equilibrium processes.  Recently,
dynamical quantities of this system such as Lyapunov exponents or
Kolmogorov-Sinai entropies in a non-equilibrium steady state have been
calculated analytically~\cite{vbdcpd,lvbd}. Systems like the Lorentz gas in
a non-equilibrium steady state can be modeled on the microscopic level by
introducing a so-called thermostat which removes the dissipated heat from
the system. Assuming that fluids can be regarded as hyperbolic systems
on a microscopic level one can relate the dynamical quantities to
viscosities or diffusion coefficients~\cite{em1990,gasp,dorfman}. This
connection is based upon a relation between phase space contraction
rates and entropy production. 

In the 1960s and 70s Sinai, Ruelle, and Bowen have developed a formalism
for dynamical system theory which due to its striking similarity to Gibbs
ensemble theory was given the name {\sl thermodynamic
formalism}~\cite{sinai,ruelle,bowen}. This formalism is applied to
hyperbolic systems and like in ordinary statistical physics a
partition function is defined which is constructed by giving points in
phase space a particular weight. From this a central quantity, the
topological pressure, is derived which is the dynamical system analog of
the Helmholtz free energy. From the dynamical partition function
properties like the Kolmogorov-Sinai (KS) entropy or the topological
entropy are obtained through the pressure and its derivative with respect
to a temperature-like parameter $\ba$. Also derivable are dimension and
entropy spectra and, for systems with escape, escape rates.

The paper is organized as follows. In Sec.~\ref{sec_tdf} we briefly
recapitulate some properties of the thermodynamic formalism.
Sec.~\ref{sec_fdl} introduces the field driven Lorentz gas and
Sec.~\ref{sec_roc} the concept of the radius of curvature. Calculations
for the field driven random Lorentz gas within the thermodynamic formalism
in two spacial dimensions are presented in Sec.~\ref{sec_calc}.  In
Sec.~\ref{sec_dyn} we calculate dynamical properties from the
thermodynamic pressure and we close with some concluding remarks in
Sec.~\ref{sec_concl}.
 
\section{Thermodynamic formalism}\label{sec_tdf}

Our starting point is the {\sl dynamical partition function}, which weighs
points in phase space by the local stretching factors
$\Lambda(\vvr,\vvv,t)$ for trajectory bundles starting at phase space
points $(\vvr,\vvv)$ and extending over a time $t$. The stretching and
contraction factors characterize the behavior of an infinitesimal volume
in phase space under the dynamics. Typically, this volume will grow in
some directions and shrink in others. Then the local stretching factor is
the factor by which the projection of the volume onto its unstable
(expanding) directions will increase over time $t$.  Similarly, the
contraction factor is the factor by which the projection of the same
volume onto its stable (contracting) direction is decreased over time $t$.
Then, the partition function is defined by 
\be
Z(\ba,t) = \int \rd\mu(\vvr,\vvv)[\Lambda(\vvr,\vvv,t)]^{1-\ba},
\label{dpf}
\ee
where a temperature-like parameter $\ba$ is introduced to emphasize the
similarity to ordinary statistical physics. The integration is over an
appropriate stationary measure.

From Eq.(\ref{dpf}) the topological pressure, $P(\ba)$, is obtained as
\be
P(\ba) = \lim_{t\rightarrow\infty}\frac{1}{t} \ln Z(\ba,t).
\label{top_p}
\ee
On introducing the Laplace transform ${\cal Z}(\ba,z) \equiv {\cal L}
\left\{ Z(\ba,t) \right\}$, the calculation of $P(\ba)$ greatly
simplifies.  From the definition of the Laplace transform we see that
${\cal Z}(\ba,z)$ only converges if $z$ stays smaller than a particular
value, the radius of convergence, which is given by
$z=\lim_{t\to\infty}\frac{1}{t}\ln Z(\ba,t)=P(\ba)$. Thus, the topological
pressure is given by the leading singularity of ${\cal
Z}(\ba,z)$,~\cite{ruelle}. 

In analogy to the standard procedures of statistical physics we can define
a {\sl dynamical entropy function}, $h(\ba)$, as the Legendre transform of
the topological pressure, i.e.
\be
h(\ba) = P(\ba) - \ba \frac{\partial P(\ba)}{\partial \ba}.
\label{entropy}
\ee

For long times the local stretching factors are approximately given by the
exponent of the sum of positive Lyapunov exponents, $\lambda_i$,
multiplied by time, $\Lambda \simeq \exp(t \sum_i^+\lambda_i)$. Therefore,
it can be shown that the entropy function defined by Eq.(\ref{entropy})
can be identified for special values of $\ba$ with dynamical properties.
For $\ba=0$, $h(\ba)$ equals the topological entropy, $h_{\rm top}$,
whereas for $\ba=1$ it equals the KS entropy, $h_{KS}$ which equals the
sum of positive Lyapunov exponents $\sum_i^+\lambda_i$, see for
instance~\cite{gasp,dorfman,beck}.

For systems where trajectories can escape, escape rates can be extracted
from the topological pressure. E.g., for $\ba=1$ the topological pressure
$P(\ba)$ equals $-\gamma$, where $\gamma$ is the escape rate while the
relationship $-\partial P(\ba)/\partial\ba|_{\ba=1} = \sum_i^+\lambda_i$
remains valid. The intersection point of $P(\ba)$ with the $\ba$-axis can
be related to the partial Hausdorff dimension, i.e.\ the fractal dimension
of a line across the stable manifold of the attractor (see~\cite{gb1995}
for details), while the intersection point of the tangent at $P(1)$ with
the $\ba$-axis is associated with the partial information
dimension~\cite{gb1995,br1987}. 

\section{Field driven Lorentz gas}\label{sec_fdl}

We study the thermodynamic formalism for the dilute, field driven Lorentz
gas without escape. This model consists of two species of particles.
Heavy, immobile particles of radius $a$ are placed at random positions,
while point particles of mass $m$ and charge $q$ move in between them. The
interaction between light and heavy particles is modeled by elastic
collisions and the heavy particles are not allowed to overlap. An external
electric field introduces a force which accelerates the moving particles
in the direction of the field and an isokinetic thermostat prevents the
system from heating up indefinitely.  This thermostat keeps the kinetic
energy and thus the speed $v$ of each moving particle
constant~\cite{em1990,hoover}. Then, the equations of motions are given by
\be
\dot\vvr = \vvv \quad , \qquad \dot\vvv = \frac{q}{m}\vve - \alpha\vvv,
\ee
where $\alpha = q\vve\cdot\vvv/mv^2 = v_x\epsilon/v$, with $\epsilon =
qE/mv$. This choice assures that the kinetic energy is kept constant. The
electric field is taken to point in the $x$-direction. 

Utilizing a Lorentz-Boltzmann equation (LBE) approach Van Beijeren {\sl et
al.} have calculated the Lyapunov spectrum of field-driven dilute Lorentz
gases in two and three spacial dimensions~\cite{vbdcpd,lvbd}.  They did so
by extending the usual LBE so as to include the so-called radius of
curvature (ROC) tensor, $\rhom$, (see Sec.~\ref{sec_roc}). The resulting
extended Lorentz-Boltzmann equation (ELBE) for the probability
distribution of $\vvr$, $\vvv$, and $\rhom$ reduces to the usual LBE by
integrating over $\rhom$. From the ELBE the KS entropy is obtained as a
steady state ensemble average.

\section{Radius of curvature}\label{sec_roc}

In order to describe the local stretching factor entering the partition
function we have to measure the separation of two nearby trajectories in
the course of time. Sinai has given geometric arguments relating this to
the ROC~\cite{sinai1970,gd1995,vbld,dorfman}.  Since the light particle
moves in $d$ dimensions, and in tangent space the direction of the flow is
neither expanding nor contracting, the unstable manifold (the subspace of
expanding directions on the energy shell in tangent space) is of dimension
$d-1$ (provided the driving field is not too strong). It may be
represented by the set of all infinitesimal displacements in position
space, orthogonal to the direction of the flow; the stretching factor for
large $t$ may be identified with the expansion factor of an infinitesimal
volume in this subspace. This expansion factor in turn may be expressed in
terms of the ROC.
   
\begin{figure}
\centerline{\includegraphics[width=0.9\columnwidth]{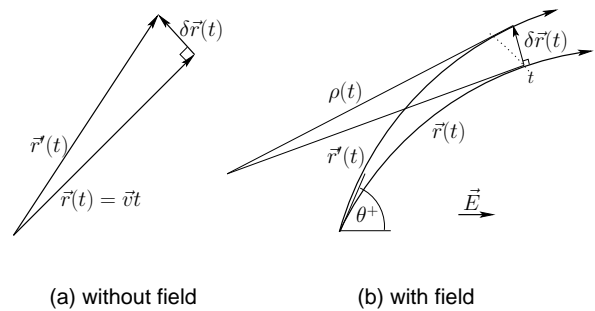}}
\caption{Illustration of the separation of two nearby trajectories,
$\vvr(t)$ and $\vvr'(t)$, for (a) no external field and (b) with an
external field, see text for details.}\label{roc}
\end{figure}
 
Fig.\ref{roc} shows a schematic plot of the radius of curvature in two
spacial dimensions (in configuration space) in the cases with and without
an external field.  We measure the separation in a plane perpendicular to
the reference trajectory. Let the spacial difference between two
trajectories be $\delta\vvr(t)$ and the difference in velocity
$\delta\vvv(t)$ after some time $t$.  These vectors satisfy the
differential equation
\be
\frac{d\delta\vvr(t)}{d t}=\delta\vvv(t).
\label{ddtdelr}
\ee
The solution of this defines the ROC tensor through the relationship
\be
\delta\vvr(t)=\frac 1 v \rhom(t,\delta\vvr(0))\cdot\delta\vvv(t).
\label{rhom}
\ee
In the absence of external fields, with $\delta\vvv$ being constant, the
ROC tensor simply is of the form $\rhom(t)=vt{\bf 1}+\rhom(0)$. Especially
in $d=2$, where $\rhom$ reduces to a scalar, $\rho(t)$ simply is the
distance of the two nearby trajectories to their mutual intersection
point, as illustrated in Fig.\ref{roc}. Its sign is positive if the
intersection is located in the past and negative if it is in the future.

One sees that already under free streaming the dependence of the ROC on
its initial conditions becomes relatively less important as time
increases. This is even much more so, if the free streaming is interrupted
by collisions.  In that case the ROC just before and just after a
collision are related in $d=2$ by
\be
\rho^{-1}_+ = \rho^{-1}_- + \frac{2}{a\cos\phi},
\label{roc_coll}
\ee
which may easily be generalized to higher dimensions~\cite{vbd,lvbd}.
Here $\rho_-$ is the pre-collisional ROC, $\rho_+$ the ROC directly after
the collision, and $\phi$ the angle between the velocity vector at the
collision and the outward normal to the scatterer at the point of
incidence. Now, the point to notice is that for dilute systems typically
$\rho^{-1}_-$ is of the order of the inverse mean free path, which is much
smaller than the second term on the right hand side of (\ref{roc_coll}).
Therefore, to leading order in the density of scatterers this term may be
ignored. This implies that the initial value of $\rho$ already gets washed
out after one collision. If one cannot use this low density approximation,
still a few collisions suffice to make the ROC independent of its initial
value. In the sequel of this paper we will use low density approximation,
so we set
\be
\rho^{-1}_+ = 2/a\cos\phi.
\label{rho+}
\ee

Combining Eqs.\ (\ref{ddtdelr}) and (\ref{rhom}) one finds that that the
stretching factor over a time $\tau$ now may be expressed in terms of the
ROC tensor as
\be
\Lambda(\tau)=\exp\left[v\int_0^{\tau} \rd t \det(\rhom^{-1}(t))\right].
\label{lambda}
\ee
From this the KS entropy may be obtained as 
\be
h_{KS} = \lim_{t\to\infty} \frac{v}{t}
\int\limits_0^t \rd \tau \det(\rhom^{-1}(\tau)) =
v\left<\det(\rhom^{-1})\right>_{\rm stat}.
\label{def_hks2d}
\ee  
The bracket $\langle \cdots \rangle_{\rm stat}$ denotes an average over a
stationary non-equilibrium distribution on phase space. Its equivalence to
a time average requires ergodicity of the motion of the moving particle on
the chaotic attractor.  About the validity of this even less is known than
about ergodicity in equilibrium, in the absence of a driving field. In our
calculations we will actually make plausible assumptions about the time
average rather than using the phase space average.

For calculating the increase of the stretching factor measured from
directly after a collision to directly after the subsequent one, one has
to integrate $\det(\rhom^{-1}(t))$ over this time interval and insert the
result into Eq.  (\ref{lambda}).  For $\cn$ uncorrelated collisions taking
place over time $t$, the corresponding stretching factor is given by the
product of the $\cn$ individual stretching factors. The KS entropy is
calculated by dividing the logarithm of the stretching factor by $t$ and
then taking the long time limit~\cite{vbld,lvbd,dorfman}.  

\section{Thermodynamic formalism for field driven Lorentz
gas}\label{sec_calc}

In~\cite{vbd} Van Beijeren and Dorfman present calculations in $d$
dimensions for the dilute random Lorentz gas without an external force.
There, the dynamical partition function is calculated by assuming that
subsequent collisions are completely uncorrelated.  Under this assumption
all free times between subsequent collisions may be assumed to be
distributed according to the same exponential function and all scattering
angles also follow the same simple distribution. In the present case,
where the moving particle still has constant speed, we will make the same
assumptions, but now the free motion in between collisions is not along
straight lines any more.

For simplicity we from now on will restrict ourselves to the case of
$d=2$, although the generalization to higher dimensions is fairly
straightforward. As in \cite{vbd} we divide up the time interval $[0,t]$
into subintervals $[t_{r-1},t_r]$, with $t_r$ the instant of the $r$-th
collision of the moving particle within this time interval. Let the total
number of collisions be $\cn$, then $t_0=0$ and $t_{\cn+1}=t$. The total
stretching factor can be factorized into a product as
\be
\Lambda(t)=\prod_{r=0}^{\cn}\Lambda_r,
\label{lambda2}
\ee
where $\Lambda_r \equiv \Lambda(\tau_r)$ is given by Eq.(\ref{lambda}),
with $\tau_r = t_{r-1} - t_r$.  For obtaining an explicit expression for
this we have to solve the differential equation describing the time
evolution of the ROC, which is of the form~\cite{vbdcpd,lvbd}
\be
\dot\rho = v + \rho\epsilon \cos\theta +
\rho^2\dfrac{\epsilon^2}{v}\sin^2\theta, 
\label{deq_rhom_2} 
\ee
where $\theta$ is the angle between the velocity vector $\vvv(t)$ and the
external field.  As initial condition we will use the low density
approximation $\rho_{r_+}=a/2\cos \phi_r$, with $\phi_r$ the collision
angle of the $r$th collision.  From the equations of motion it follows
that $\theta$ obeys the differential equation $\dot\theta = -\epsilon
\sin\theta$ which has the solution
\be
\theta(t,\theta^+) = \arccot\Big[\sinh\big(\epsilon t + \kappa \big)\Big], \quad
\kappa = \ln\big[\cot(\theta^+/2)\big]. 
\label{sol_theta}
\ee
Here, $\theta^+=\theta(0,\theta^+_r)$ is the angle between the external
field and the velocity direction directly after the $r$th collision,
compare Fig.\ref{roc}(b). With this solution for $\theta(t)$ we rewrite
Eq.(\ref{deq_rhom_2}) as
\be 
\dot\rho = v + \rho\epsilon\tanh(\epsilon t + \kappa) +
\rho^2\frac{\epsilon^2}{v \cosh^2(\epsilon t + \kappa)}. 
\label{deq_rho}
\ee 
We find for the ROC
\be
\rho(t) 
= \frac{v}{\epsilon}\cosh(\epsilon t +
\kappa)\frac{\sinh(\epsilon t +\kappa) -
\frac{v\cosh\kappa\sinh\kappa - \epsilon\rho_+}{v\cosh \kappa +
\epsilon\rho_+\sinh\kappa}}{1+\sinh(\epsilon t
+\kappa)
\frac{v\cosh\kappa\sinh\kappa -
\epsilon\rho_+}{v\cosh\kappa
+ \epsilon\rho_+\sinh\kappa}}. \label{sol_rho2d}
\ee
Inserting this into Eq.(\ref{lambda}) we find, to second order in $\epsilon$, 
\bea
\Lambda_r &=& \Lambda_r^{(1)}\Lambda_r^{(2)} \label{split}\\
\Lambda_r^{(1)} &=& \frac{v\tau_r}{\rho_{r_+}} \label{lambda(1)}\\
\Lambda_r^{(2)} = \exp\Big[-\frac{\epsilon\tau_r}{2}\cos\theta^+_{r}
&+&
\frac{\epsilon^2\tau_r^2}{24}\left(1-9\sin^2\theta^+_{r}\right) 
\Big].\quad 
\label{stretch2d}
\eea
Note that the above equations may also be used for $r=0$ but with
$\rho_{_{0_+}}$ and $\theta^+_{0}$ given by the initial values of $\rho$
respectively $\theta$ at $t=0$.

For obtaining the topological pressure one has to substitute the above
results into Eq.(\ref{lambda2}), raise this to the power $(1-\beta)$ and
average over the configurations of scatterers. For simplifying this
average it turns out to be useful to rearrange Eq.(\ref{lambda2}) as
\be
\Lambda(t)=v\tau_{_\cn}\Lambda_{_\cn}^{(2)} \left[\prod_{r=1}^{\cn}
\Gamma_{_{\cn-r}}\right]
\ \rho_{_{0_+}}^{-1}
\ee
where
\be
\Gamma_r = \Gamma (\tau_r, \theta^+_{r+1}, \theta^+_r) \equiv
\frac{v\tau_r}{\rho_{_{(r+1)_+}}} \Lambda^{(2)}_r .
\label{lambdat2}
\ee
The reduction to the independent variables implied here follows from
Eq.(\ref{rho+}) combined with the relationship
\be
2\phi_{r}=\theta^+_{r}-\theta^-_{r}\pm \pi
\ee
and $\theta^-_{r}$ may be expressed in terms of $\theta^+_{r-1}$ and
$\tau_{r-1}$ through Eq.(\ref{sol_theta}).  
 
A further simplification can be made by passing to the Laplace transform
${\cal{Z}}(\ba,z)$ of the dynamical partition function. Since, in the
limit of large $t$, the topological pressure equals the logarithm of the
stretching factor per unit time, it may be identified as the rightmost
singularity of this Laplace transform. It has to be real, as the
stretching factor is real and positive definite.  Assuming independence of
all free flight times $\tau_j$ and scattering angles $\phi_{j}$ one finds
straightforwardly that ${\cal{Z}}(\ba,z)$ may be obtained as
\bea
{\cal{Z}}(\ba,z)&=&\sum\limits_{\cn=0}^\infty\int
\rd\theta^+_{\cn}\cdots \int \rd\theta^+_{0} \ \hat{\cal M}_f (\ba,z,
\theta^+_{\cn})
\nonumber\\
&& \times 
\prod\limits_{r=1}^{\cn}\hat{{\cal
M}}(\ba,z,\theta^+_{\cn-r+1},\theta^+_{\cn-r})
\ \rho_{_{0_+}}^{-1}
\nonumber\\
&=&
\sum\limits_{\cn=0}^\infty \left\{\hat{\mathfrak M}_f
\hat{\mathfrak
M}^\cn
\ \rho_{_{0_+}}^{-1}
\right\}(\ba,z,\cdot)
\nonumber\\
&=&
\left\{\hat{\mathfrak M}
_f
\left[{\bf1}-\hat{\mathfrak M}\right]^{-1}
\ \rho_{_{0_+}}^{-1}\right\}(\ba,z,\cdot) \quad
\label{calz}
\eea
Here the operators $\hat{{\cal M}}_f$ and $\hat{{\cal M}}$ are defined as
the Laplace transforms of the configurational averages of the appropriate
powers of $v\tau_\cn\Lambda_\cn^{(2)}$ and $\Gamma(\tau_r, \theta_{r+1},
\theta_r)$, respectively. Specifically, one has
\bea
\hat{{\cal M}}(\ba,z,\theta^+{'},\theta^+)&=&\int\limits_0^{\infty}
\rd\tau \ \nu \ e^{-(z+\nu)\tau}\int\limits^{\pi/2}_{-\pi/2} \rd\phi \
\frac{\cos\phi}{2} \nonumber \\
&&
\times \ \delta\left[\phi -
\frac{1}{2}\left(\theta^+{'}-\theta(\tau,\theta^+)\pm\pi\right)\right] 
\label{operators}
\nonumber \\ 
&&
\times \ \left[\Gamma(\tau,
\theta^+{'},\theta^+)\right]^{1-\ba} , \\
\hat{\cal M}_f(\ba,z,
\theta^+_{\cn})&=&
V\int\limits_0^{\infty} \rd\tau_\cn \
\nu \ e^{-(z+\nu)\tau_\cn}
\left(v\tau_\cn\Lambda_\cn^{(2)}\right)^{1-\ba} \quad
\eea
with $\theta(\tau,\theta^+)$ defined in Eq.(\ref{sol_theta}). The
operators
$\hat{\mathfrak M}$
and  $\hat{\mathfrak M}_f$
are defined by
\bea
\left\{\hat{\mathfrak M}f\right\}(\ba,z,\theta^+{'}) &=& \int \rd\theta^+
\hat{\cal M} (\ba,z,\theta^+{'},\theta^+) f( \theta^+)\\
\left\{\hat{\mathfrak M}_f f\right\}(\ba,z,\cdot) &=& \int \rd\theta^+
\hat{\cal M}_f (\ba,z, \theta^+)f(\theta^+)
\eea
where integration limits of the $\theta^+ $-integrations depend on the
sign of $\pi$ in the delta function in Eq.(\ref{operators}) and are
given in the Appendix.  Further, ${\bf 1}$ denotes the unit operator and
$[\cdots ]^{-1}$ the operator inverse.

From the above equations the rightmost singularity of ${\cal{Z}}$ readily
follows as the value of $z$ for which the largest eigenvalue of the
operator $\hat{{\mathfrak M}}$ equals unity. For zero field this
eigenvalue problem is trivial: the leading eigenfunction is the unit
function, the eigenvalue is obtained easily and the resulting topological
pressure coincides with that found in~\cite{vbd}. For small non-zero field
one has to proceed by expanding $\hat{{\mathfrak M}}$, the leading
eigenfunction $f$ and the leading eigenvalue $\mu$ in powers of the field
strength, i.e.\ 
\bea
\hat{\mathfrak M} &=& \hat{\mathfrak M}^{(0)} +
\epsilon 
\hat{\mathfrak M}^{(1)} + \epsilon^2 \hat{\mathfrak M}^{(2)}
\label{transopexp} \\
f &=& f^{(0)} + \epsilon f^{(1)} +\epsilon^2
f^{(2)} \\
\mu &=& \mu^{(0)} + \epsilon \mu^{(1)} + \epsilon^2
\mu^{(2)}.
\eea
Then $\hat{\mathfrak M} f = \mu f$ is solved by standard
perturbation methods using a Fourier series expansion for $f$, i.e.\
\be
f^{(i)} = a_o^{(i)} + \sum_{n=1}^{\infty} a_n^{(i)}
\cos(n\theta^+) + \sum_{n=1}^{\infty} b_n^{(i)} \sin(n\theta^+),
\label{fourier}
\ee
with $i=0, 1, 2$.

The details of the solution of this eigenvalue equation are given in the
Appendix~\ref{appa}. Here we just give the resulting eigenvalue $\mu$ to
second order in the field strength,
\bea
\mu(\ba,z) &=& \mu^{(0)}(\ba,z) + \epsilon^2 \mu^{(2)}(\ba,z) \nonumber
\\
&=& \nu\left(\frac{v}{a}\right)^{1-\ba}
\frac{\Gamma\left(\half\right)\Gamma\left(\frac{\ba+1}{2}\right)\Gamma(2-\ba)}{2^\ba\Gamma\left(\frac{\ba}{2}+1\right)(\n+z)^{2-\ba}}
\nonumber \\
&& \times \Bigg[1 +
\epsilon^2 \frac{(\ba-1)(\ba-2)(\ba-12)}{48(\nu+z)^2}
\Bigg]. \quad
\label{lap_z2d}
\eea
Now the Laplace transform of the dynamical partition function is of the
form
\be
{\cal Z} (\ba,z) = 
C(\ba,z) \left[ 1 - \mu(\ba,z) \right]^{-1},
\ee
where the additional prefactor $C$, originating from $\hat{\mathfrak
M}(\ba,z,\cdot,\theta^+_{\cn})$, contains no singularities.

\begin{figure}
\centerline{\includegraphics[width=0.8\columnwidth]{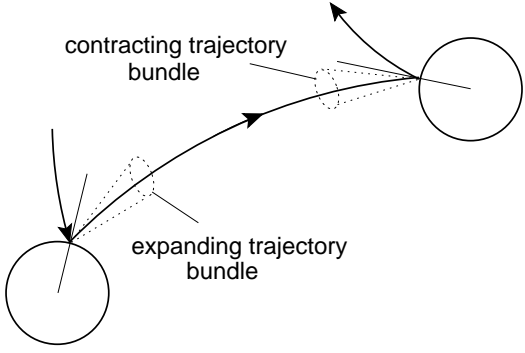}}
\caption{Illustration of expanding and contracting trajectory bundles for
the two dimensional field driven random Lorentz gas.
}\label{roc_traj}
\end{figure}

If we also want to obtain information about the contracting direction,
e.g.\ the negative Lyapunov exponents, we need to know the local
contraction factors. For this we may consider the time reversed motion.
Note, that we will consider the time reversed motion on the attractor for
the ``forward in time motion'' and not the attractor of the time reversed
motion. The contraction factors can be calculated by considering
contracting trajectory bundles instead of expanding ones~\cite{vbld}, see
Fig.\ref{roc_traj}. 
We keep the same notation as for the forward in time motion, therefore
collision $r$ precedes collision $r-1$ in the time reversed motion.  Hence
the boundary condition for $\rho(t)$ in Eq.(\ref{deq_rho}) is
$\rho(\tau)=\rho_-=a\cos\phi /2$, where $\rho_-$ specifies the ROC
directly before a collision (in the forward in time motion). The ROC still
evolves in time according to Eq.(\ref{deq_rho}) for $0\leq t\leq\tau$.
Like the expansion factors, one calculates the local contraction factors
by using Eq.(\ref{lambda}).  To second order in $\epsilon$ this yields
\bea
\Lambda^-_r &=& \Lambda_r^{(1)-}\Lambda_r^{(2)-}, \label{split-}\\
\Lambda_r^{(1)} &=& \frac{\rho_{r_-}}{v\tau_r}, \label{lambda(1)-}\\
\Lambda_r^{(2)} = \exp\Big[-\frac{\epsilon\tau_r}{2}\cos\theta^+_{r}
&+& 
\frac{\epsilon^2\tau_r^2}{24}\left(1+3\sin^2\theta^+_{r}\right) 
\Big], \qquad
\label{stretch2dneg}
\eea
with $\rho_{r_{-}} = a \cos\phi_r/2$.  In the limit of vanishing external
field the contraction factor is just the inverse of the stretching factor.
This is no longer true if we apply an external field. We see from
Eqs.\ (\ref{stretch2d}) and (\ref{stretch2dneg}) that differences occur in
the field dependent exponential.

To obtain a topological pressure for the contracting directions we have to
solve a similar eigenvalue problem as for the expanding directions. 
Since for the field free case the inverse of the contraction factor equals
the stretching factor, we solve the eigenvalue problem for the inverse of
the contraction factor. 
The resulting eigenvalue, which has to be set equal to unity again, to
second order in field strength becomes
\bea
\mu^-(\ba,z)
&=&
\nu\left(\frac{v}{a}\right)^{1-\ba}
\frac{\Gamma\left(\half\right)\Gamma\left(\frac{\ba+1}{2}\right)\Gamma(2-\ba)}{2^\ba\Gamma\left(\frac{\ba}{2}+1\right)(\n+z)^{2-\ba}}
\nonumber \\
&& \times \Big[1 -
\epsilon^2 \frac{(\ba-1)(\ba-2)(\ba-8)}{48(\nu+z)^2}
\Big]\label{lap_z2d_neg}.
\eea
Details are given again in the Appendix~\ref{appa}. 

In the following section we will discuss the resulting topological pressure
and related properties.

\section{Dynamical properties}\label{sec_dyn}

As stated in the preceding section the topological pressure follows as the
value of $z$ for which $\mu(\ba,z)=1$. To second order in $\epsilon$ this
leads to
\be
P(\ba)= P_0(\ba) - \epsilon^2 \frac{(\ba-1)(\ba-12)}{48(P_0(\ba)+\nu)}
\ee
where
\be
P_0(\ba) = \left[ \frac{\nu}{2}
\Big(\frac{v}{a}\Big)^{(1-\ba)} \
\frac{\Gamma(2-\ba)\Gamma\left(\half\right)\Gamma\left(\frac{\ba+1}{2}\right)}{2^\ba
\Gamma\left(\frac{\ba}{2}+1\right)} \right]^{\frac{1}{2-\ba}} - \nu
\ee
is the field free value of the topological pressure.
The dynamical entropy then follows from Eq.(\ref{entropy}) as
\bea
h(\ba) &=& P(\ba) - \ba \frac{\partial P(\ba)}{\partial \ba} \nonumber \\
&=& h_0(\ba) - \frac{\epsilon^2}{48(P_0(\ba)+\nu)} \Bigg[ 12-\ba^2
\nonumber \\
&& + \frac{\ba(\ba-1)(\ba-12)}{P_0(\ba)+\nu} \cdot \frac{\partial
P_0(\ba)}{\partial \ba} \Bigg]
\label{top_entr_result}
\eea
where again $h_0(\ba)$ is the field free value of the dynamical entropy.

We can perform similar calculations for the contracting direction. Then we
obtain the equivalent of the topological pressure and the dynamical
entropy. Note that in the limit of vanishing external field the
topological pressure for the contracting direction equals the one for the
expanding direction. We obtain
\bea
P^-(\ba) &=& P_0(\ba) +
\epsilon^2\frac{(\ba-1)(\ba-8)}{48(P_0(\ba)+\nu)} \\
h^-(\ba) &=& h_0(\ba) + \frac{\epsilon^2}{48(P_0(\ba)+\nu)} \Bigg[ 8 -
\ba^2 \nonumber \\
&& + \frac{\ba(\ba-1)(\ba-8)}{P_0(\ba)+\nu}\cdot
\frac{\partial
P_0(\ba)}{\partial \ba} \Bigg] .
\label{top_entr_result_neg}
\eea

\begin{figure}[b]
\centerline{\includegraphics[clip=,width=0.8\columnwidth]{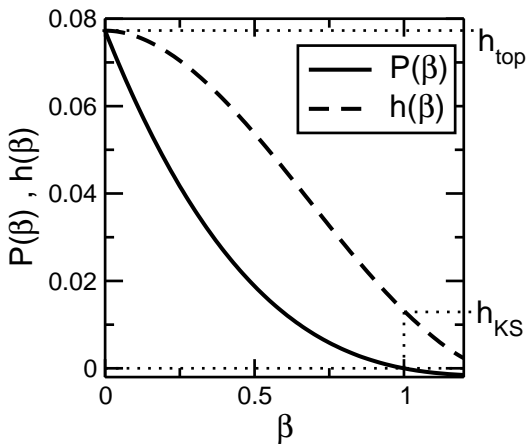}}
\caption{The topological pressure, $P(\ba)$, and
the {\sl dynamical} entropy, $h(\ba)$, for small fields and small
densities of scatterers, i.e. $\epsilon=0.001, \ n=0.001, \ v=1, \ a=1$.
}\label{ruelle_press}
\end{figure}

Fig.\ref{ruelle_press} shows the topological pressure and the dynamical
entropy as functions of the parameter $\ba$. As expected for a system
without escape, $P(\ba)$ vanishes for $\ba=1$. We further see that
$P(\ba)$ is a convex function. 

A number of interesting dynamical quantities are related to the dynamical
entropy, $h(\ba)$, for special values of $\ba$. The KS entropy, $h_{KS}$,
is given by the dynamical entropy for $\ba=1$.  From
Eq.(\ref{top_entr_result}) we obtain
\be
h_{KS} = \lambda^+ = \n\left(1-\gamma-\ln\left(\frac{a\n}{v}\right)\right) -
\frac{11}{48}\frac{\epsilon^2}{\n}.
\ee
For the negative Lyapunov exponent we yield, from
Eq.(\ref{top_entr_result_neg})
\be
-h^-(1) = \lambda^- =
-\n\left(1-\gamma-\ln\left(\frac{a\n}{v}\right)\right) -
\frac{7}{48}\frac{\epsilon^2}{\n}. 
\ee

As one should expect, these results coincide with those of previous
calculations~\cite{vbdcpd,lvbd} based on the same assumptions (basically,
no correlations between subsequent collisions).  But here the results are
extended to general values of $\ba$.

New results follow for the topological entropy, which is given by the
dynamical entropy for $\ba=0$. From Eq.(\ref{top_entr_result}) 
we obtain 
\be
h_{\rm top} = \left[\frac{\pi\nu v}{a}\right]^{\half} - \nu 
- \frac{\epsilon^2}{4} \left[\frac{a}{\pi\nu v}\right]^{\half}
\label{htop2d}
\ee
That is, we obtain the zero field limit results given in~\cite{vbd} with a
correction which is quadratic in the field strength. 

The equivalend of the topological entropy for the contracting direction is
obtained from Eq.(\ref{top_entr_result_neg}) for $\ba=0$, i.e.
\be
h^-(0) = \left[\frac{\pi\nu v}{a}\right]^{\half} -\nu + \frac{\epsilon^2}{6}
\left[\frac{a}{\pi\nu v}\right]^{\half}.
\ee

\begin{figure}
\centerline{\includegraphics[clip=,width=0.8\columnwidth]{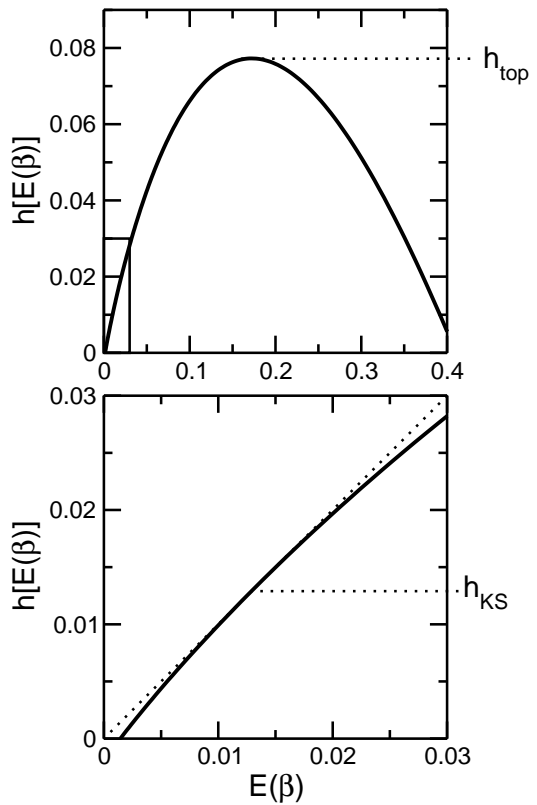}}
\caption{Dynamical entropy, $h[E(\ba)]$, as a function of
$E(\ba)=-\partial P(\ba)/\partial\ba$ for $\epsilon=0.001$, $n=0.001$,
$a=1$, and $v=1$. The lower panel is a close-up of the boxed region of the
upper panel.}\label{entr_eng}
\end{figure}

\begin{figure}
\centerline{\includegraphics[clip=,width=0.75\columnwidth]{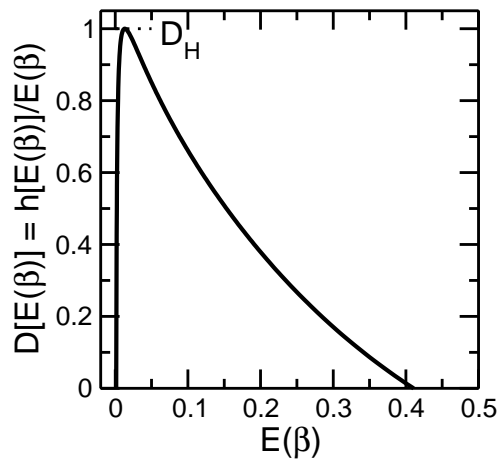}}
\caption{Dimension spectrum for $\epsilon=0.001$, $n=0.001$, $a=1$, and
$v=1$. The maximum occurs for $\ba=1$.}\label{dimspec}
\end{figure}

We further calculate the dynamical entropy as a function of
$E(\ba)\equiv-\partial P(\ba)/\partial\ba$,~\cite{br1987}. $E(\ba)$ equals
the average of the logarithm of the local stretching factors,
$\left<\ln\Lambda\right>_\ba$, where the subscript $\ba$ refers to the
fact that in phase space initial points are weighted according to the
stretching factors raised to the power $(1-\ba)$. For $\ba=1$ this yields
the average of the positive Lyapunov exponent.

The maximum entropy, which is the topological entropy, is always found for
$\ba=0$ as can be seen from Eq.(\ref{entropy}).  The KS entropy is given
by the value for $\ba=1$ where $h[E(\ba)]=E(\ba)$, again according to
Eq.(\ref{entropy}). In Fig.\ref{entr_eng} a plot of $h[E(\ba)]$ vs.
$E(\ba)$ is given. But notice that the descending parts of $h[E(\ba])$
correspond to values of $\ba\leq0$. 

The dynamical entropy is related to a dimension spectrum, $D(\alpha)$,
with $\alpha=E(\ba)$, through $D(\alpha)=h(\alpha)/\alpha$ \cite{br1987}.
Fig.\ref{dimspec} shows the dimension spectrum, $D[E(\ba)]$. From this one
can find the partial Hausdorff dimension, $D_H$, as the maximum of
$D[E(\ba)]$. For systems where trajectories cannot escape, $D_H$ equals
one.  This is clear because the maximum of $D[E(\ba)]$ is obtained for
$P(\ba)=0$, i.e.\ $\ba=1$, for systems without escape. Note that possible
values of partial dimensions are restricted to the interval $[0,1]$, see
Sec.\ref{sec_tdf}. Furthermore, the full Hausdorff dimension is $D_H^{\rm
full} \leq \sum_{i=1}^ 3 D_H^{(i)}$ where the $D_H^{(i)}$ are the partial
Hausdorff dimensions corresponding to all stable and unstable
manifolds~\cite{er1985}.

The above results for the dynamical entropy allow to calculate,
respectively approximate, another dimension, the Kaplan-Yorke dimension,
$D_{KY}$. In general the Kaplan-Yorke dimension is given by $D_{KY} = n +
\sum_{i=1}^n \lambda_i /|\lambda_{n+1}|$ where $n$ is given by the largest
value for which $\sum_{i=1}^n \lambda_i \geq 0$. Thus for the two
dimensional Lorentz gas with constant energy we have one exponent equal to
$0$, one positive, and one negative, i.e.\ we have $D_{KY} = 2 + \lambda^+
/|\lambda^-|$. There are only three exponents since the system is
restricted to a three dimensional hyperplane by the thermostat.  From the
KS entropies we see that $D_{KY} \approx 3 - 3\epsilon^2/(8 h_{KS}^0
\nu)$, where $h_{KS}^0$ is the KS entropy without external field.  Since
the exact full Hausdorff dimension and therefore also the full information
dimension cannot be extracted from the partial dimensions we assume that
the Kaplan-Yorke conjecture still holds.  This is consistent with our
present knowledge: because $D_H^{\rm full} \leq \sum_{i=1}^3 D_H^{(i)} =
3$ we still can have $D_H^{\rm full} = D_{KY}$.  Let us further mention
that the dimensional loss due to the external field, as expressed by
$D_{KY}$, is rather small in the region where the above results hold, i.e.
for small fields and low densities, see also~\cite{ecsb}. 

The calculated dynamical properties allow for the extraction of
macroscopic transport coefficients.  The diffusion coefficient is given by
${\cal D}=-\lim_{\epsilon\to0}[v^2(\lambda_+(\epsilon) +
\lambda_-(\epsilon))]\epsilon^{-2} = (3/8)v^2/\nu$,~\cite{vbdcpd}, which
can also be expressed in terms of the Kaplan-Yorke dimension by ${\cal
D}=-\lim_{\epsilon\to0} [v^2 h^0_{KS}(D_{KY}-
3)]\epsilon^{-2}$,~\cite{ecsb}.

Some comment on the calculations of the topological pressure as a function
of the temperature-like parameter $\ba$ is in order here. The obtained
results have to be taken with a pinch of salt. In~\cite{vbd}
and~\cite{avbed} it is shown that for the random Lorentz gas the results
obtained there are restricted to increasingly smaller neighborhoods of
$\ba\approx1$ for increasing system size. As can be seen from
Eq.(\ref{dpf}), for $\beta=1$ all points in phase space are equally
weighted.
For $\ba<1$, though, the partition function will be dominated by the
largest stretching factors, which correspond to the most unstable
trajectory bundles.  That is, for $\ba<1$ stretching factors from regions
in phase space with a high density of scatterers, and therefore large
stretching factors, dominate.  So, in the limit $t\to\infty$ it is
possible that $Z(\ba,t)$ is dominated by trajectory bundles confined to a
small part of phase space. In regions of high scatterer density however,
subsequent scattering events cannot be regarded as independent any more
and the distribution of free times between scatterers in these regions
will be very different from that for the system as a whole. With
increasing system size the effects become more pronounced because the
probability of finding approximately trapping regions of high scatterer
density increases with system size.

\section{Conclusion}\label{sec_concl}

In the present study we have calculated dynamical properties for the field
driven random Lorentz gas within the thermodynamic formalism. In the
limits of vanishing external field or $\ba $ approaching unity, our
results are in perfect agreement with those of previous studies. 

From the topological pressure we extracted various quantities, such as the
KS entropy and the topological entropy. A dimension spectrum was obtained
by calculating the dynamical entropy as a function of the variable
$E(\ba)$, defined as the derivative with respect to $\ba$ of the
topological pressure. 

Van Beijeren and Dorfman have calculated KS entropies and topological
entropies for general dimensions $d$ for the random Lorentz gas without an
external field~\cite{vbd}. Presently we work on the extension of the
present study to higher dimensions. Subtleties exist because an external
field complicates the analytic calculation of the determinant of the
inverse of the ROC tensor for higher dimensions quite a bit. However, the
stretching factors can also be calculated by looking at the time evolution
of the deviations in velocity, see~\cite{vbld}.

We also extended our studies to systems with open
boundaries~\cite{mvbopen}. This allows for using the thermodynamic
formalism to study the escape rate formalism~\cite{gn1990}, as well as
dimension spectra for systems with escape. In the limit of $\ba\to 1$
comparison can be made again with previous results~\cite{vbld2000}.

\begin{acknowledgments}
We thank Bob Dorfman for helpful discussions and valuable comments.  This
work was supported by the {\sl Collective and cooperative statistical
physics phenomena} program of FOM (Fundamenteel Onderzoek der Materie).
\end{acknowledgments}

\appendix*
\section{Calculation of the dynamical partition function}\label{appa}

Starting with the dynamical partition function we can determine the
topological pressure which is given as the leading singularity of the
Laplace transform of the dynamical partition function. 

In order to calculate the Laplace transform of the dynamical partition
function, Eq.(\ref{calz}), we need to calculate the eigenfunctions and
eigenvalues of $\hat{\mathfrak M}$.  According to Eq.(\ref{transopexp}) we
expand $\hat{\mathfrak M}$ in powers of $\epsilon$, yielding
\bw
\bea
\hat{\mathfrak M}(\ba,z,\theta^+{'}) 
&=& \int\limits_{0}^{\infty} \rd\tau \int\limits_{-\pi/2}^{\pi/2} \rd\phi
\int \rd\theta^+
\ \nu \ e^{-(\nu+z)\tau}
\ \frac{\cos\phi}{2} 
\ \delta\left[\phi - \frac{1}{2}(\theta^+{'} -
\theta(\tau,\theta^+) \pm \pi
\right]\left[\frac{2v\tau}{a\cos\phi}\right]^{1-\ba}  \nonumber \\
&& \times \left\{ 1 -
\epsilon \frac{\tau}{2} \cos\theta^+ + \epsilon^2 
\frac{\tau^2}{24} \left[ 1 - 9\sin\theta^+ + 3\cos^2\theta^+
\right] \right\}^{1-\ba},
\label{operators_app}
\eea
The operator $\hat{\mathfrak M}$ has to be understood as acting on a
function $f(\theta^+)$.  In order to eliminate the $\delta$-function we
integrate over $\theta^+$ first by noticing that we can write
$\delta[g(\theta^+)] = \delta(\theta^+ - \theta^+_0)/g'(\theta^+_0)$ where
$\theta^+_0$ is the root of $g(\theta^+_0)$ and the prime denotes the
derivative with respect to $\theta^+$.  Here we have
\be
g'(\theta^+_0) = \half - \frac{\epsilon\tau}{2} \cos\theta^+_0 +
\frac{(\epsilon\tau)^2}{4} \cos(2\theta^+_0)
\ee
with 
\be
\theta^+_0 = \theta^+{'} - 2\phi +\epsilon\tau \sin\theta^+_0 -
\frac{(\epsilon\tau)^2}{4} \sin(2\theta^+_0) \pm \pi.
\ee
Then the operator $\hat{\mathfrak M}(\ba,z,\theta^+{'})$ becomes 
\bea
\hat{\mathfrak M}(\ba,z,\theta^+{'})
&=& \int\limits_{0}^{\infty} \rd\tau \int\limits_{-\pi/2}^{\pi/2} \rd\phi 
\ \nu \ e^{-(\nu+z)\tau}
\ \frac{\cos\phi}{2} \left[\frac{2v\tau}{a\cos\phi}\right]^{1-\ba}
2 \left( 1 + \epsilon\tau \cos\theta^+_0 +
\frac{(\epsilon\tau)^2}{2} \right)
\nonumber \\
&& \times 
\left\{ 1 -
\epsilon \frac{\tau}{2} \cos\theta^+ + \epsilon^2 
\frac{\tau^2}{24} \left[ -2 + 6\cos(2\theta^+)
\right] \right\}^{1-\ba} 
\eea
This is expanded to second order and the operator $\hat{\mathfrak M}$
acting on $f(\theta^+)$ then reads 
\be
\hat{\mathfrak M} f(\theta^+) = \left[\hat{\mathfrak M} f \right]
(\theta^+{'}) = \left[\hat{\mathfrak M}^{(0)} f \right] (\theta^+{'}) +
\epsilon \left[\hat{\mathfrak M}^{(1)} f \right] (\theta^+{'}) +
\left[\epsilon^2 \hat{\mathfrak M}^{(2)} f \right] (\theta^+{'})
\label{operators_app_theta}
\ee

\bea
\mbox{with} \quad \left[\hat{\mathfrak M}^{(0)} f \right] (\theta^+{'}) 
&=& 
\int\limits_{0}^{\infty} \rd\tau \int\limits_{-\pi/2}^{\pi/2} \rd\phi 
\ \nu \ e^{-(\nu+z)\tau}
\ \left[\frac{2v\tau}{a}\right]^{1-\ba} \cos^\ba\phi \ f(\theta^+_0)
\\
\left[\hat{\mathfrak M}^{(1)} f \right] (\theta^+{'})
&=& 
- \int\limits_{0}^{\infty} \rd\tau \int\limits_{-\pi/2}^{\pi/2} \rd\phi 
\ \nu \ e^{-(\nu+z)\tau}
\ \left[\frac{2v\tau}{a}\right]^{1-\ba} \cos^\ba\phi \ \frac{\tau}{2}
(1+\ba) \cos\theta^+_0 \ f(\theta^+_0)
\label{exp_op_1}
\\
\left[\hat{\mathfrak M}^{(2)} f \right] (\theta^+{'})
&=& 
\int\limits_{0}^{\infty} \rd\tau \int\limits_{-\pi/2}^{\pi/2} \rd\phi 
\ \nu \ e^{-(\nu+z)\tau}
\ \left[\frac{2v\tau}{a}\right]^{1-\ba} \cos^\ba\phi 
\nonumber \\
&& \quad \times \frac{\tau^2}{2}\left(
\frac{3\ba^2+13\ba+8}{24} + \frac{\ba(\ba-1)}{8} \cos(2\theta^+_0) \right)
\ f(\theta^+_0)
\eea
\ew
For small $\epsilon$ we can expand $\cos\theta^+_0$ and
$\cos(2\theta^+_0)$ which gives
\bea
\cos\theta^+_0 &=& \cos[\theta^+{'} - 2\phi \pm \pi + \epsilon\tau
\sin(\theta^+{'} - 2\phi \pm \pi)] \nonumber \\
&\approx& - \cos(\theta^+{'} - 2\phi) - \epsilon\tau \sin^2(\theta^+{'} -
2\phi)
\label{cos_theta}
\eea
and
\be
\cos(2\theta^+_0) \approx \cos(2\theta^+{'} - 4\phi).
\ee
Since we expand only to second order in $\epsilon$ it is sufficient to
expand $\cos\theta^+_0$ to first order because it only enters in the
$\epsilon$-term of Eq.(\ref{operators_app_theta}). Accordingly, we only
take the zeroth order term of $\cos(2\theta^+_0)$. 
Now the eigenvalues $\mu$ and eigenfunctions $f$ can be calculated by
using standard perturbation theory and a Fourier series expansion for $f$
as given in Eq.(\ref{fourier}).  To zeroth order in $\epsilon$ we find
that $f^{(0)}(\theta^+) = {\rm constant }$, which we set equal to $1$. For
the eigenvalue we get
\be \mu^{(0)} = \nu \Big(\frac{v}{a}\Big)^{(1-\ba)} \
\frac{\Gamma(2-\ba)\Gamma\left(\half\right)\Gamma\left(\frac{\ba+1}{2}\right)}{2^\ba(\nu+z)^{2-\ba}\Gamma\left(\frac{\ba}{2}+1\right)}.
\ee
Inserting these results into Eq.(\ref{exp_op_1}) we get
\bea
\mu^{(1)} &=& 0 \\
\mbox{and} \quad f^{(1)}(\theta^+) &=&
- \frac{\ba(2-\ba)}{4(\nu+z)}\cos\theta^+.
\eea

The procedure for the $\epsilon^2$-terms is analogous. However, when
calculating the eigenvalue $\mu^{(2)}$ one finds an additional contribution
from $\left[\hat{\mathfrak M}^{(0)} f^{(1)}\right] (\theta^+{'})$, which
also has to be taken into account. This is because of the
$\epsilon$-dependent term in Eq.(\ref{cos_theta}).
Then to second order in $\epsilon$ the eigenvalue is given by
\be
\mu^{(2)} 
= \mu^{(0)} \frac{(\ba-1)(\ba-2)(\ba-12)}{48(\nu+z)^2}
\ee
In principle the eigenfunction $f^{(2)}(\theta^+)$ can be calculated and
will be proportional to $\cos(2\theta^+)$. However, for our results we
do not need $f^{(2)}(\theta^+)$ since the only terms entering in
Eq.(\ref{lap_z2d}) are $\mu^{(0)} f^{(0)}(\theta^+)$ and $\epsilon^2
\mu^{(2)} f^{(0)}(\theta^+)$.

An analogous calculation for the contracting direction yields for the
eigenvalues
\bea
\bar\mu^{(0)} &=& \nu \Big(\frac{v}{a}\Big)^{(1-\ba)} \
\frac{\Gamma(2-\ba)\Gamma\left(\half\right)\Gamma\left(\frac{\ba+1}{2}\right)}{2^\ba(\nu+z)^{2-\ba}\Gamma\left(\frac{\ba}{2}+1\right)}
\\
\bar\mu^{(1)} &=& 0\\
\bar\mu^{(2)} &=& - \bar\mu^{(0)}
\frac{(\ba-1)(\ba-2)(\ba-8)}{48(\nu+z)^2}
\eea
and for the eigenfunctions
\bea
\bar f^{(0)}(\theta^+) &=& 1 \\
\bar f^{(1)}(\theta^+) &=& 
- \frac{\ba(2-\ba)}{4(\nu+z)}\cos\theta^+,
\eea
where the bar is indicating the contracting direction. Again, for
Eq.(\ref{lap_z2d_neg}) we only need the eigenfunctions up to first order in
$\epsilon$.


\begin{thebibliography}{22}
\expandafter\ifx\csname natexlab\endcsname\relax\def\natexlab#1{#1}\fi
\expandafter\ifx\csname bibnamefont\endcsname\relax
  \def\bibnamefont#1{#1}\fi
\expandafter\ifx\csname bibfnamefont\endcsname\relax
  \def\bibfnamefont#1{#1}\fi
\expandafter\ifx\csname citenamefont\endcsname\relax
  \def\citenamefont#1{#1}\fi
\expandafter\ifx\csname url\endcsname\relax
  \def\url#1{\texttt{#1}}\fi
\expandafter\ifx\csname urlprefix\endcsname\relax\def\urlprefix{URL }\fi
\providecommand{\bibinfo}[2]{#2}
\providecommand{\eprint}[2][]{\url{#2}}

\bibitem[{\citenamefont{van Beijeren et~al.}(1996)\citenamefont{van Beijeren,
  Dorfman, Cohen, Posch, and Dellago}}]{vbdcpd}
\bibinfo{author}{\bibfnamefont{H.}~\bibnamefont{van Beijeren}},
  \bibinfo{author}{\bibfnamefont{J.~R.} \bibnamefont{Dorfman}},
  \bibinfo{author}{\bibfnamefont{E.~G.~D.} \bibnamefont{Cohen}},
  \bibinfo{author}{\bibfnamefont{H.~A.} \bibnamefont{Posch}}, \bibnamefont{and}
  \bibinfo{author}{\bibfnamefont{C.}~\bibnamefont{Dellago}},
  \bibinfo{journal}{Phys.\ Rev.\ Lett.} \textbf{\bibinfo{volume}{77}},
  \bibinfo{pages}{1974} (\bibinfo{year}{1996}).

\bibitem[{\citenamefont{Latz et~al.}(1997)\citenamefont{Latz, van Beijeren, and
  Dorfman}}]{lvbd}
\bibinfo{author}{\bibfnamefont{A.}~\bibnamefont{Latz}},
  \bibinfo{author}{\bibfnamefont{H.}~\bibnamefont{van Beijeren}},
  \bibnamefont{and} \bibinfo{author}{\bibfnamefont{J.~R.}
  \bibnamefont{Dorfman}}, \bibinfo{journal}{Phys.\ Rev.\ Lett.}
  \textbf{\bibinfo{volume}{78}}, \bibinfo{pages}{207} (\bibinfo{year}{1997}).

\bibitem[{\citenamefont{Evans and Morriss}(1990)}]{em1990}
\bibinfo{author}{\bibfnamefont{D.~J.} \bibnamefont{Evans}} \bibnamefont{and}
  \bibinfo{author}{\bibfnamefont{G.~P.} \bibnamefont{Morriss}},
  \emph{\bibinfo{title}{Statistical Mechanics of Nonequilibrium Liquids}}
  (\bibinfo{publisher}{Academic Press, London}, \bibinfo{year}{1990}).

\bibitem[{\citenamefont{Gaspard}(1998)}]{gasp}
\bibinfo{author}{\bibfnamefont{P.}~\bibnamefont{Gaspard}},
  \emph{\bibinfo{title}{Chaos, Scattering, and Statistical Mechanics}}
  (\bibinfo{publisher}{Cambridge University Press, Cambridge UK},
  \bibinfo{year}{1998}).

\bibitem[{\citenamefont{Dorfman}(1999)}]{dorfman}
\bibinfo{author}{\bibfnamefont{J.~R.} \bibnamefont{Dorfman}},
  \emph{\bibinfo{title}{An Introduction to Chaos in Non-Equilibrium Statistical
  Mechanics}} (\bibinfo{publisher}{Cambridge University Press, New York},
  \bibinfo{year}{1999}).

\bibitem[{\citenamefont{Sinai}(1972)}]{sinai}
\bibinfo{author}{\bibfnamefont{Y.~G.} \bibnamefont{Sinai}},
  \bibinfo{journal}{Russ.\ Math.\ Surv.} \textbf{\bibinfo{volume}{27}},
  \bibinfo{pages}{21} (\bibinfo{year}{1972}).

\bibitem[{\citenamefont{Ruelle}(1978)}]{ruelle}
\bibinfo{author}{\bibfnamefont{D.}~\bibnamefont{Ruelle}},
  \emph{\bibinfo{title}{Thermodynamic Formalism}}
  (\bibinfo{publisher}{Addison-Wesley Publishing Co., New York},
  \bibinfo{year}{1978}).

\bibitem[{\citenamefont{Bowen}(1975)}]{bowen}
\bibinfo{author}{\bibfnamefont{R.}~\bibnamefont{Bowen}}, in
  \emph{\bibinfo{booktitle}{Lecture Notes in Mathematics}}
  (\bibinfo{publisher}{Springer Verlag, Berlin}, \bibinfo{year}{1975}), vol.
  \bibinfo{volume}{470}.

\bibitem[{\citenamefont{Beck and Schl{\"o}gl}(1993)}]{beck}
\bibinfo{author}{\bibfnamefont{C.}~\bibnamefont{Beck}} \bibnamefont{and}
  \bibinfo{author}{\bibfnamefont{F.}~\bibnamefont{Schl{\"o}gl}},
  \emph{\bibinfo{title}{Thermodynamics of chaotic systems}}
  (\bibinfo{publisher}{Cambridge University Press, New York},
  \bibinfo{year}{1993}).

\bibitem[{\citenamefont{Gaspard and Baras}(1995)}]{gb1995}
\bibinfo{author}{\bibfnamefont{P.}~\bibnamefont{Gaspard}} \bibnamefont{and}
  \bibinfo{author}{\bibfnamefont{F.}~\bibnamefont{Baras}},
  \bibinfo{journal}{Phys.\ Rev.\ E} \textbf{\bibinfo{volume}{51}},
  \bibinfo{pages}{5332} (\bibinfo{year}{1995}).

\bibitem[{\citenamefont{Bohr and Rand}(1987)}]{br1987}
\bibinfo{author}{\bibfnamefont{T.}~\bibnamefont{Bohr}} \bibnamefont{and}
  \bibinfo{author}{\bibfnamefont{D.}~\bibnamefont{Rand}},
  \bibinfo{journal}{Physica D} \textbf{\bibinfo{volume}{25}},
  \bibinfo{pages}{387} (\bibinfo{year}{1987}).

\bibitem[{\citenamefont{Hoover}(1991)}]{hoover}
\bibinfo{author}{\bibfnamefont{W.~G.} \bibnamefont{Hoover}},
  \emph{\bibinfo{title}{Computational Statistical Mechanics}}
  (\bibinfo{publisher}{Elsevier Publ.\ Co., Amsterdam}, \bibinfo{year}{1991}).

\bibitem[{\citenamefont{Gaspard and Dorfman}(1995)}]{gd1995}
\bibinfo{author}{\bibfnamefont{P.}~\bibnamefont{Gaspard}} \bibnamefont{and}
  \bibinfo{author}{\bibfnamefont{J.~R.} \bibnamefont{Dorfman}},
  \bibinfo{journal}{Phys.\ Rev.\ E} \textbf{\bibinfo{volume}{52}},
  \bibinfo{pages}{3525} (\bibinfo{year}{1995}).

\bibitem[{\citenamefont{van Beijeren et~al.}(1998)\citenamefont{van Beijeren,
  Latz, and Dorfman}}]{vbld}
\bibinfo{author}{\bibfnamefont{H.}~\bibnamefont{van Beijeren}},
  \bibinfo{author}{\bibfnamefont{A.}~\bibnamefont{Latz}}, \bibnamefont{and}
  \bibinfo{author}{\bibfnamefont{J.~R.} \bibnamefont{Dorfman}},
  \bibinfo{journal}{Phys.\ Rev.\ E} \textbf{\bibinfo{volume}{57}},
  \bibinfo{pages}{4077} (\bibinfo{year}{1998}).

\bibitem[{\citenamefont{Sinai}(1970)}]{sinai1970}
\bibinfo{author}{\bibfnamefont{Y.~G.} \bibnamefont{Sinai}},
  \bibinfo{journal}{Russ.\ Math.\ Surv.} \textbf{\bibinfo{volume}{25}},
  \bibinfo{pages}{137} (\bibinfo{year}{1970}).

\bibitem[{\citenamefont{van Beijeren and Dorfman}(2002)}]{vbd}
\bibinfo{author}{\bibfnamefont{H.}~\bibnamefont{van Beijeren}}
  \bibnamefont{and} \bibinfo{author}{\bibfnamefont{J.~R.}
  \bibnamefont{Dorfman}}, \bibinfo{journal}{J.\ Stat.\ Phys.}
  \textbf{\bibinfo{volume}{108}}, \bibinfo{pages}{767} (\bibinfo{year}{2002}).

\bibitem[{\citenamefont{van Beijeren et~al.}(2000)\citenamefont{van Beijeren,
  Latz, and Dorfman}}]{vbld2000}
\bibinfo{author}{\bibfnamefont{H.}~\bibnamefont{van Beijeren}},
  \bibinfo{author}{\bibfnamefont{A.}~\bibnamefont{Latz}}, \bibnamefont{and}
  \bibinfo{author}{\bibfnamefont{J.~R.} \bibnamefont{Dorfman}},
  \bibinfo{journal}{Phys.\ Rev.\ E} \textbf{\bibinfo{volume}{63}},
  \bibinfo{pages}{016312} (\bibinfo{year}{2000}).

\bibitem[{\citenamefont{Eckmann and Ruelle}(1985)}]{er1985}
\bibinfo{author}{\bibfnamefont{J.-P.} \bibnamefont{Eckmann}} \bibnamefont{and}
  \bibinfo{author}{\bibfnamefont{D.}~\bibnamefont{Ruelle}},
  \bibinfo{journal}{Rev.\ Mod.\ Phys.} \textbf{\bibinfo{volume}{57}},
  \bibinfo{pages}{617} (\bibinfo{year}{1985}).

\bibitem[{\citenamefont{Evans et~al.}(2000)\citenamefont{Evans, Cohen, Searles,
  and Bonetto}}]{ecsb}
\bibinfo{author}{\bibfnamefont{D.~J.} \bibnamefont{Evans}},
  \bibinfo{author}{\bibfnamefont{E.~G.~D.} \bibnamefont{Cohen}},
  \bibinfo{author}{\bibfnamefont{D.~J.} \bibnamefont{Searles}},
  \bibnamefont{and} \bibinfo{author}{\bibfnamefont{F.}~\bibnamefont{Bonetto}},
  \bibinfo{journal}{J.\ Stat.\ Phys.} \textbf{\bibinfo{volume}{101}},
  \bibinfo{pages}{17} (\bibinfo{year}{2000}).

\bibitem[{\citenamefont{Appert et~al.}(1996)\citenamefont{Appert, van Beijeren,
  Ernst, and Dorfman}}]{avbed}
\bibinfo{author}{\bibfnamefont{C.}~\bibnamefont{Appert}},
  \bibinfo{author}{\bibfnamefont{H.}~\bibnamefont{van Beijeren}},
  \bibinfo{author}{\bibfnamefont{M.~H.} \bibnamefont{Ernst}}, \bibnamefont{and}
  \bibinfo{author}{\bibfnamefont{J.~R.} \bibnamefont{Dorfman}},
  \bibinfo{journal}{Phys.\ Rev.\ Lett.} \textbf{\bibinfo{volume}{54}},
  \bibinfo{pages}{54} (\bibinfo{year}{1996}).

\bibitem[{\citenamefont{M{\"u}lken and van Beijeren}(2003)}]{mvbopen}
\bibinfo{author}{\bibfnamefont{O.}~\bibnamefont{M{\"u}lken}}
\bibnamefont{and}
  \bibinfo{author}{\bibfnamefont{H.}~\bibnamefont{van Beijeren}},
  \bibinfo{journal}{unpublished}.

\bibitem[{\citenamefont{Gaspard and Nicolis}(1990)}]{gn1990}
\bibinfo{author}{\bibfnamefont{P.}~\bibnamefont{Gaspard}} \bibnamefont{and}
  \bibinfo{author}{\bibfnamefont{G.}~\bibnamefont{Nicolis}},
  \bibinfo{journal}{Phys.\ Rev.\ Lett.} \textbf{\bibinfo{volume}{65}},
  \bibinfo{pages}{1693} (\bibinfo{year}{1990}).

\end{thebibliography}
\end{document}